# Magnetocaloric properties of single-crystalline Eu$_5$In$_2$Sb$_6$

Karol Synoradzki[1,*], Tomasz Toliński[1], Qurat Ul Ain[1], Michał Matczak[1], Tetiana Romanova[2], Dariusz. Kaczorowski[2]

[1] Institute of Molecular Physics, Polish Academy of Sciences, Poznań, Poland
[2] Institute of Low Temperature and Structure Research, Polish Academy of Sciences, Wrocław, Poland

## ABSTRACT

The magnetocaloric effect (MCE) in single-crystalline Eu$_5$In$_2$Sb$_6$ was studied in broad ranges of temperature and magnetic field intervals, with the magnetic field oriented parallel and perpendicular to the crystallographic $c$-axis in the orthorhombic Ca$_5$Ga$_2$As$_6$-type unit cell of the crystals. The compound is an antiferromagnetic insulator with two subsequent magnetic phase transitions at $T_{N1} \approx 14.0$ K and $T_{N2} \approx 7.1$ K. The compound was found to show both normal and inverse magnetocaloric effect (MCE). Furthermore, owing to substantial magnetic anisotropy, it exhibits also a rotational MCE. For conventional MCE, the maximum value of isothermal entropy change $\Delta S_m$ is -4.7(1) J kg$^{-1}$ K$^{-1}$ at 15(1) K for a magnetic field change of 5 T, independent of the field direction. The relative cooling power (RCP) is also almost identical for both directions, and equals to 81(1) J kg$^{-1}$ for a magnetic field change by 5 T. The most significant rotational MCE was observed at about 11 K, with maximum $\Delta S_R$ value of 2.4(1) J kg$^{-1}$ K$^{-1}$ and RCP = 17.5(6) J kg$^{-1}$ for a magnetic field change of 4 T.

Keywords: Eu$_5$In$_2$Sb$_6$, magnetocaloric effect, rotational magnetocaloric effect,

* Corresponding author
Email address: karol.synoradzki@ifmpan.poznan.pl (K. Synoradzki)

K. Synoradzki: orcid 0000-0002-7732-3284

**INTRODUCTION**

Europium-based compounds have recently gained much attention, due to their unique optical, magnetic, electronic and quantum properties, which can be utilized in wide range of applications, from lighting and displays to quantum information processing and medical imaging [1]. Especially interesting are materials bearing divalent Eu ions, which are magnetic and often order into complex magnetic structures at low temperatures.

The compound $Eu_5In_2Sb_6$ is a semiconducting Zintl phase, which crystallizes in the orthorhombic $Ca_5Ga_2As_6$-type structure (space group *Pbam*, no. 55) with nonsymorphic symmetry elements [2]. The compound is an antiferromagnet (AFM) with two subsequent second-order magnetic phase transitions at $T_{N1} \approx$ 14.0 K and $T_{N2} \approx$ 7.1 K [3–9]. Spectroscopic measurements indicated that all the Eu ions in its crystallographic unit cell are divalent [4,9]. In the paramagnetic state, $Eu_5In_2Sb_6$ exhibits colossal magnetoresistance (CMR) being as large as -99.999% at 15 K in a field of 9 T [4], which is one of the highest values observed for stoichiometric antiferromagnets. Interestingly, the compound shows also a large and strongly anisotropic piezoresistance [10]. Application of uniaxial pressure along the [001] direction by a mere 0.4 GPa culminates in a remarkable reduction in the resistivity of over 99.95%, which yields a colossal piezoresistance factor of 5000 $\times 10^{-11}$ Pa$^{-1}$. The immense susceptibility of the electronic transport to external stimuli may render $Eu_5In_2Sb_6$ a valuable candidate for the detection of dark matter [11].

As a narrow-gap semiconductor, $Eu_5In_2Sb_6$ has garnered attention for its potential as a thermoelectric (TE) material. While the TE performance in pure $Eu_5In_2Sb_6$ is not substantial, and strongly sample dependent [2,12,13], it can be significantly enhanced through proper modifications of the chemical composition. For example, upon substitution of 10% Cd or 5% Zn for In, the dimensionless TE figure of merit (*zT*) increases to 0.4-0.5 observed at temperatures near 700 K [12,13].

In this study, we attempted to determine the magnetocaloric properties of $Eu_5In_2Sb_6$. Our work was motivated by the unique physical properties of the compound, but also the observation that several Eu-based materials exhibit large values of the magnetocaloric parameters such as the magnetic entropy change $\Delta S_m$ or the adiabatic temperature change $\Delta T_{ad}$. A good example is EuSe for which $\Delta S_m$ reaches almost 40 J kg$^{-1}$ K$^{-1}$ at about 5 K with a magnetic field change of 5 T [14], which is one of the highest values seen in any material in this temperature range, and opens a prospective of using it in cryogenic magnetic refrigerators.

**MATERIALS & METHODS**

Single crystals of $Eu_5In_2Sb_6$ were grown using the indium flux method [15]. Pure elements (Eu - 99.9 at.%, In - 99.999 at.%, and Sb - 99.999 at.%), taken in a ratio 3:110:6, were placed in an ACP-CCS-5 Canfield Crucible Set (LCP Industrial Ceramics Inc.) and sealed in an evacuated quartz ampoule. The tube was slowly heated up to 1000 °C, kept at that temperature for 24 hours, and then cooled to 500 °C with a rate of 2 °C/hour. Subsequently, the flux was removed by centrifuging. The so-obtained crystals were dark grey and had a form of flat rods with dimensions up to 1 mm in the *ab* plane and a few mm along the *c* axis.

X-ray diffraction (XRD) measurements were performed at room temperature on a powder prepared from several hand-ground crystals using a PANalytical X'pert Pro diffractometer with CuKα-radiation, generated at 40 kV and 30 mA ($\lambda$ = 1.5406 Å), in a Bragg-Brentano geometry. The XRD patterns were analyzed using FullProf software [16]. The crystal density was measured by the Archimedes method at room temperature using isopropyl alcohol and an electronic balance. The chemical composition was examined employing a FEI Technologies scanning electron microscope (SEM) equipped with a Genesis XM4 energy dispersive x-ray spectrometer (EDS).

Magnetization measurements were carried out on the oriented single crystals in the temperature range 1.8 - 300 K in magnetic fields up to 5 T using a Quantum Design MPMS-XL superconducting quantum interference device (SQUID) magnetometer, and in the temperature range 2 - 300 K in magnetic fields up to 9 T employing a Quantum Design Physical Property Measuring System (PPMS-9) equipped with a vibrating sample magnetometer (VSM). The specific heat was measured from 2 to 295 K in magnetic fields up to 9 T using the relaxation technique and the two-τ method implemented in the PPMS apparatus [17].

**RESULTS & DISCUSSION**

*Crystal-chemical characterization*

The obtained XRD data confirmed that the obtained crystals of $Eu_5In_2Sb_6$ have an orthorhombic crystal structure of the $Ca_5Ga_2As_6$-type (space group of *Pbam*, oP26, No. 55). The Eu atoms occupy three different Wyckoff positions: Eu1 and Eu3: 4$g$ ($x$, $y$, 0) and Eu2: 2a (0, 0, 0). The Sb atoms are also located in three Wyckoff positions: Sb1: 4$g$ ($x$, $y$, 0); Sb2 and Sb3: 4$h$ ($x$, $y$, 1/2). In turn, the In atoms occupy only one Wyckoff site 4$h$ ($x$, $y$, 1/2). The Rieveld refinement of the XRD pattern (see Fig. 1a) yielded the lattice parameters: $a$ = 12.516(1) Å, $b$ = 14.598(2) Å, $c$ = 4.633(1) Å, which are in good agreement with values reported in the literature. The experimentally determined density of 6.73(1) g cm$^{-3}$ is only slightly smaller than that calculated from the XRD data (6.748 g cm$^{-3}$), which corroborates good quality of the crystals examined.

As can be inferred from the SEM image shown Fig. 1b, the obtained single crystals were homogeneous and free of imperfections. The EDS analyses conducted on several crystal surfaces revealed an average chemical composition of 4.9(1):2.0(1):6.1(1), closely matching the nominal stoichiometry 5:2:6. Further proof of good crystalline quality of the samples examined was obtained from the EDS composition mapping, visualized in Fig. 1(c-e). The results revealed uniform distribution of all the elements across the sample surface, except for spots of indium flux residues (seen also on the XRD pattern; cf. Fig. 1a).

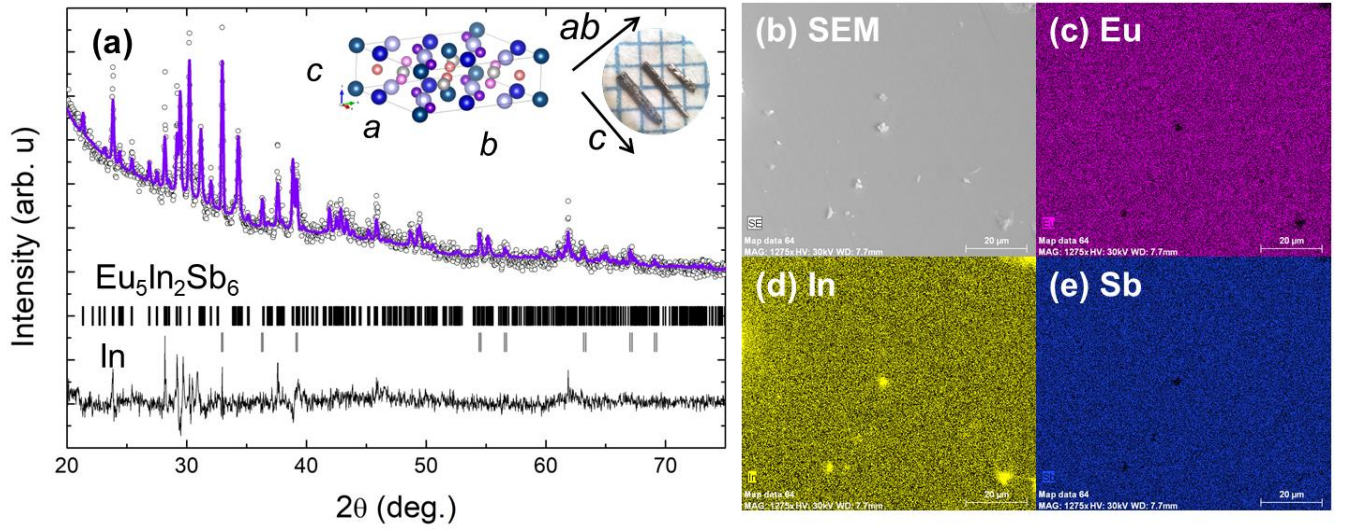

Fig. 1. Crystal-chemical properties of $Eu_5In_2Sb_6$. (a) Powder X-ray diffraction pattern. The circles represent the experimental data, the solid line is the Rietveld fit. The difference between the experimental results and the fit is shown by the solid line at the bottom. The upper row of vertical ticks corresponds to the Bragg peaks positions of the compound, while the lower one refers to metallic In flux remnants. The left inset shows the orthorhombic unit cell of the compound. The right inset shows a photo of the obtained single crystals; (b) SEM image; (c,d,e) EDS maps obtained for individual elements: Eu, In, and Sb, respectively.

*Magnetic properties*

The results of magnetic measurements are collected in Figure 2. $Eu_5In_2Sb_6$ exhibits two subsequent magnetic phase transitions of antiferromagnetic character, one at $T_{N1}$ = 14.7(1) K and another one at $T_{N2}$ = 7.6(1) K. The values of these temperatures were determined as inflections in the $M(T)$ curves measured in a magnetic field of 0.1 T. They are very similar to those reported before [4–7,12]. The zero-field-cooled (ZFC) and field-cooled (FC) curves overlap (not shown), hence indicating the absence of any ferromagnetic component.

In the paramagnetic state, the inverse magnetic susceptibility is essentially independent of the magnetic field direction, as expected for a spin-only divalent Eu ion. It can be well fitted with the Curie–Weiss (CW) law:

$$\chi^{-1}(T) = \left[\frac{N_A \mu_{eff}^2}{3k_B(T-\theta_{CW})}\right]^{-1}, \qquad (1)$$

where $N_A$ is the Avogadro number, $\mu_{eff}$ is the effective magnetic moment, $k_B$ is the Boltzmann constant, and $\theta_{CW}$ is the CW temperature. The CW parameters obtained from the fit are: $\mu_{eff}$ = 7.98(1) $\mu_B$/Eu, $\theta_{CW}$ = -1.7(1) K, for $H \parallel c$, and $\mu_{eff}$ = 7.92(1) $\mu_B$/Eu, $\theta_{CW}$ = 0.4(3) K, for $H \parallel ab$. The different signs of $\theta_{CW}$ reflect fairly complex magnetic structure of the compound, governed by predominant antiferromagnetic interactions between the Eu magnetic moments along the $c$-axis, and ferromagnetic interactions within the $ab$-plane [8,9]. The $\mu_{eff}$ values are very close to the theoretical Russel-Saunders prediction for $Eu^{2+}$ ion: ($g\sqrt{J(J+1)}$ = 7.94; where $g$ is the Lande factor and $J$ is the total angular momentum).

In the ordered state, the magnetic properties of $Eu_5In_2Sb_6$ are strongly anisotropic, as observed in the temperature dependencies of the magnetic susceptibility (see the upper inset to Figure 2), and the field dependencies of the magnetization, measured at 2 K (see the lower inset to Figure 2). The $M/H(T)$ curves taken in $H \parallel ab$ and $H \parallel c$ distinctly differ from each other in respect of both their shapes and amplitudes. The $M(H)$ dependence measured with $H \parallel c$ is almost linear and featureless, while that obtained with $H \parallel ab$ shows a sharp metamagnetic transition near $\mu_0 H = 2.4$ T. Regardless the magnetic field direction no saturation in $M(H)$ is attained up to 9 T. Generally, the magnetic behavior of the crystals of $Eu_5In_2Sb_6$ studied in the present work agrees very well with the previous reports [4,7,2].

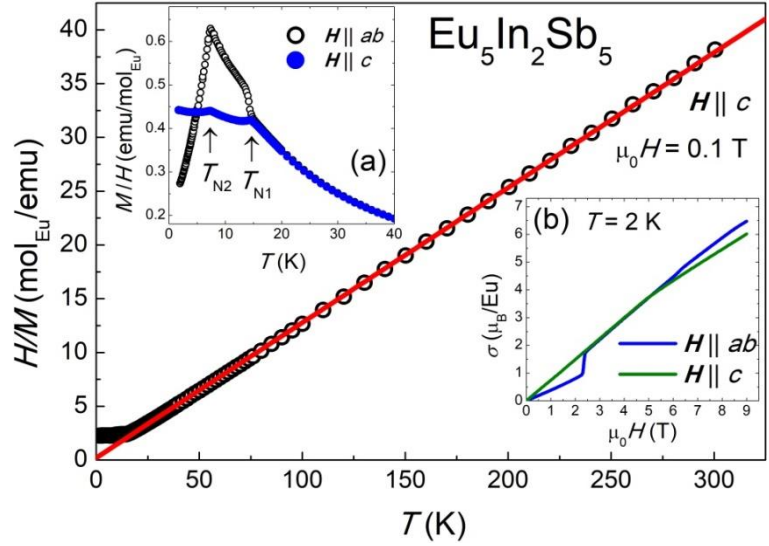

Fig. 2. Temperature dependence of the inverse magnetic susceptibility of single-crystalline $Eu_5In_2Sb_6$ measured in a magnetic field of 0.1 T applied along the $c$-axis. Solid red line presents a fit with the CW law. (a) Temperature variations of the magnetic susceptibility of $Eu_5In_2Sb_6$ in the $ab$-plane and along the $c$-axis. (b) Magnetic field dependencies of the magnetization of single-crystalline $Eu_5In_2Sb_6$ taken at $T = 2$ K in rising fields oriented in the $ab$-plane and along the $c$-axis.

*Magnetocaloric properties*

Figures 3(a) and 3(b) show the isothermal dependencies of $M(H)$ measured at several temperatures in the vicinity of the magnetic transitions in $Eu_5In_2Sb_6$ in magnetic field applied within the $ab$-plane and along the $c$-axis, respectively. The same isotherms are presented in Figures 3(c) and 3(d) in the form of Arrott plots. A clear dissimilarity between the data collected in different field orientations is mostly a consequence of the presence of the first-order metamagnetic transition along the $ab$-plane that in absent in the other direction.

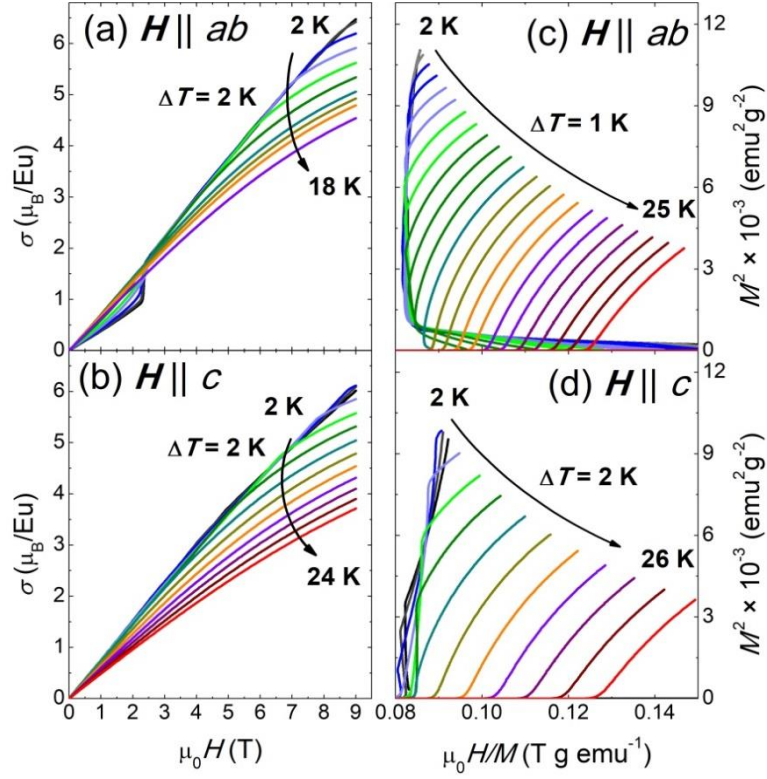

Fig. 3. The magnetization isotherms measured for single-crystalline $Eu_5In_2Sb_6$ with (a) $H \parallel ab$ and (c) $H \parallel c$, shown also as Arrott plots in panels (c) and (d), respectively.

Employing the Maxwell relation [18]:

$$\Delta S_m(T,H) = \int_0^H (\delta M/\delta T) dH , \qquad (2)$$

the isothermal entropy change $\Delta S_m$ in $Eu_5In_2Sb_6$ was calculated. As can be inferred from Figure 4, for both magnetic field orientations, a prevalence of normal magnetocaloric effect (MCE) with a negative value of $\Delta S_m$ was found. The $\Delta S_m(T)$ curves have a shape of an asymmetric peak, which is steeper on the lower temperature side, with a maximum slightly above $T_{N1}$. The maximum value of the entropy change ($\Delta S_m^{MAX}$) for the magnetic field changes of 5 T is -4.7(1) J kg$^{-1}$ K$^{-1}$ for both $H \parallel ab$ and $H \parallel c$ directions. For $H \parallel a$, an inverse MCE with the maximum magnetic entropy change equal to 2.2(1) J kg$^{-1}$ K$^{-1}$ was obtained near 12 K for $\mu_0 \Delta H = 3$ T.

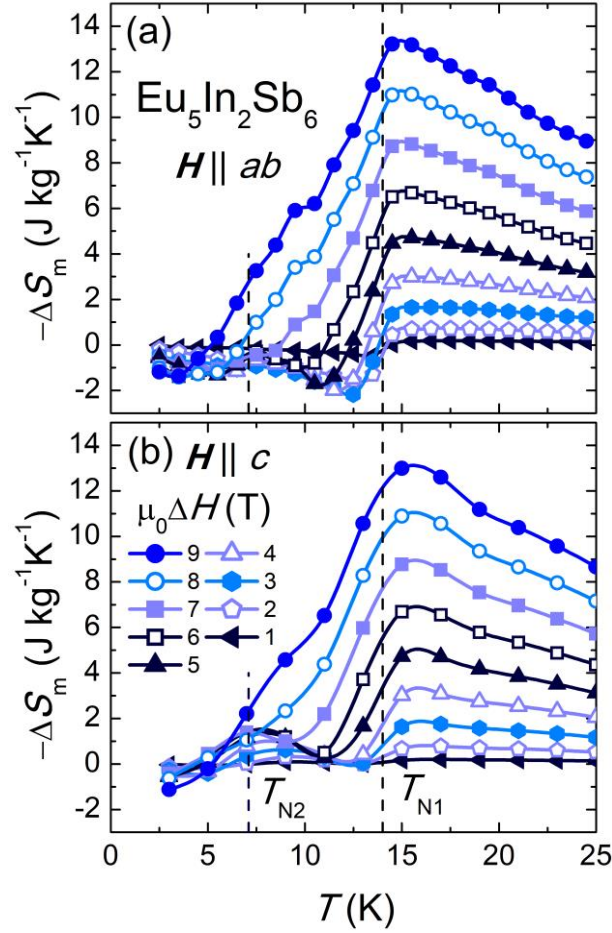

Fig. 4. Temperature dependencies of the magnetic entropy change in single-crystalline $Eu_5In_2Sb_6$ determined for different values of the magnetic field change $\Delta H$ for (a) $\mathbf{H} \parallel ab$ and (b) $\mathbf{H} \parallel c$. The solid lines serve as a guide for the eye. The vertical dashed lines represent the AFM magnetic phase transition temperatures $T_{N1}$ and $T_{N2}$ observed in the magnetic measurements.

Figure 5 shows the isothermal entropy change in $Eu_5In_2Sb_6$ as a function of $\Delta H$, derived for $H \parallel ab$ and $H \parallel c$ for a few selected temperatures. In the paramagnetic state, a power-law behavior $\Delta S_m \sim \Delta H^n$ was found with the critical exponent of about 1.7 that differs from the values predicted by theoretical models for AFM systems [19]. On the other hand, the estimated value of $n$ is reasonably close to $n = 2$, characteristic of ferromagnets above $T_C$, as can be obtained by integrating the CW law [20].

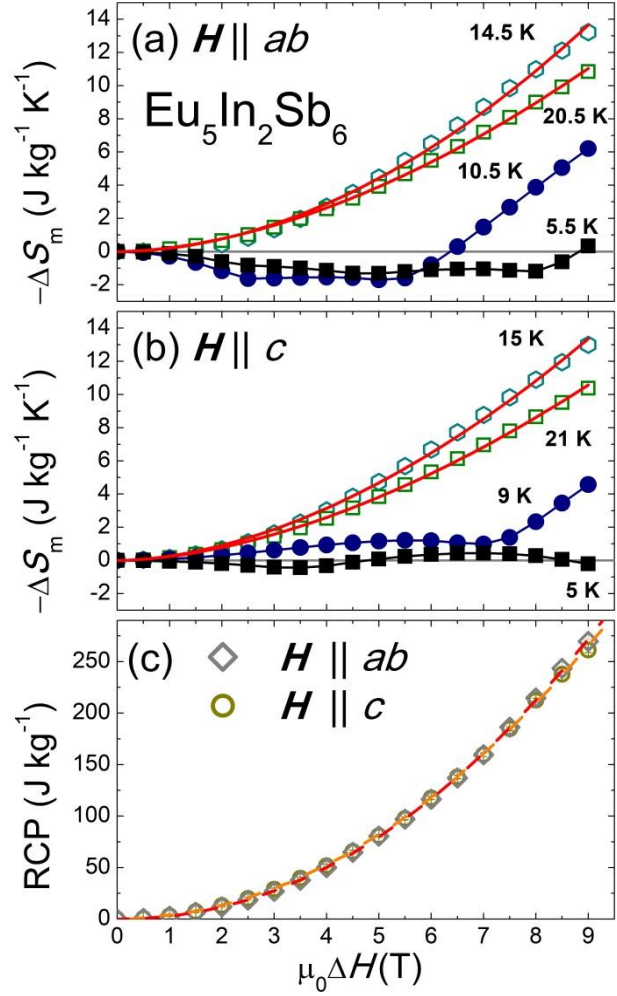

Fig. 5. The magnetic entropy change as a function of the magnetic field change, estimated for single-crystalline $Eu_5In_2Sb_6$ at selected temperatures for (a) $H \parallel ab$ and (b) $H \parallel c$. The red solid lines represent the power law fits of the data measured in the paramagnetic state. (c) The relative cooling power in single-crystalline $Eu_5In_2Sb_6$ with conventional MCE as a function of the magnetic field change, determined for the $H \parallel ab$ and $H \parallel c$ orientations. The dashed lines represent the power low fits, described in the text.

An important parameter, commonly employed to assess the thermocaloric performance of materials, is the relative cooling power (RCP) that reflects heat exchange capability between hot and cold thermal reservoirs. The magnitude of RCP can be estimated from the formula [21]:

$$\text{RCP} = |\Delta S_m^{\text{MAX}}| \delta T_{\text{FWHM}} , \qquad (3)$$

where $\delta T_{\text{FWHM}}$ stands for the full-width-at-half-maximum of the $\Delta S_m(T)$ peak. In the case of $Eu_5In_2Sb_6$, the maximum RCP value for magnetic field change of 5 T is almost the same for both magnetic field directions, and equals 81(1) J kg$^{-1}$ (see Fig. 5(c)). If the magnetic field change is increased to 9 T, the RCP value increases to 270(2) J kg$^{-1}$. The rise of RCP can usually be described by a power law dependence RCP ~ $\Delta H^m$ [20]. For $Eu_5In_2Sb_6$, the exponent $m$ somehow deviates from the values 1.2 - 1.333, predicted by classical models [22].

When assessing the potential performance of magnetocaloric materials, it proves also beneficial to compute the temperature-averaged entropy change (TEC) [23]. This parameter can be derived from the formula [23]:

$$\text{TEC}(\Delta T_{lift}) = \frac{1}{\Delta T_{lift}} \max_{T_{mid}} \left\{ \int_{T_{mid}-\Delta T_{lift}/2}^{T_{mid}+\Delta T_{lift}/2} \Delta S_m(T)\, dT \right\}, \quad (4)$$

where $\Delta T_{\text{lift}}$ denotes the desired temperature span of the device for a given $\Delta H$, and $T_{\text{mid}}$ represents the averaged temperature. In this study of $Eu_5In_2Sb_6$, the temperature interval $\Delta T_{\text{lift}} = 5$ K was chosen to calculate the field dependence of TEC(5) for normal MCE. For $\boldsymbol{H} \parallel ab$, TEC(5) = 0.68(1), 4.5(1), and 12.6(1) J kg$^{-1}$ K$^{-1}$ were obtained for a magnetic field change of 2, 5, and 9 T, respectively. In turn, TEC(5) = 0.78(1), 4.5(1), and 12.6(1) J kg$^{-1}$ K$^{-1}$ respectively, were derive for the same $\Delta H$ if $\boldsymbol{H} \parallel c$. These values are comparable to the maximum values of $\Delta S_m$ for the given values of the magnetic field change.

Single crystals of magnetic materials have the potential to be used in rotating magnetocaloric devices, where the refrigeration force is created by rotating an anisotropic sample in a constant magnetic field instead of inducing a phase transition in an isotropic material by changing magnetic field [24]. The rotational magnetocaloric effect (RMCE) induces a positive adiabatic temperature change when shifting the magnetic field from hard to easy direction of magnetization. In materials designed for such purposes, the primary focus lies in maximizing the difference between values observed along two distinct crystallographic directions. A larger difference between these values corresponds to an enhanced refrigerant capacity when the sample undergoes rotation in magnetic field. The rotational magnetic entropy change $\Delta S_R$ value is related to magnetocrystalline anisotropy, and can be determined as the difference between the magnetic entropy change values determined for two different crystallographic directions :

$$\Delta S_R(T, H) = \Delta S_m(T, H_{ab}) - \Delta S_m(T, H_c) \, . \quad (5)$$

Figure 6(a) shows the $\Delta S_R(T)$, $\Delta S_m(T, H_{ab})$ and $\Delta S_m(T, H_c)$ data derived for $Eu_5In_2Sb_6$ for a magnetic field change of 4 T, at which the largest values of $\Delta S_R$ were recorded. The results of $\Delta S_R(T)$ calculations for different values of magnetic field changes are shown in Figure 6(b), and Figure 6(c) presents RCP estimated for RMCE. Though the RCP magnitude is rather small, and does not exceed 20 J/kg, it should be noted that it initially increases with increasing $\Delta H$ and for $\mu_0 \Delta H > 4$ T, it start to decrease.

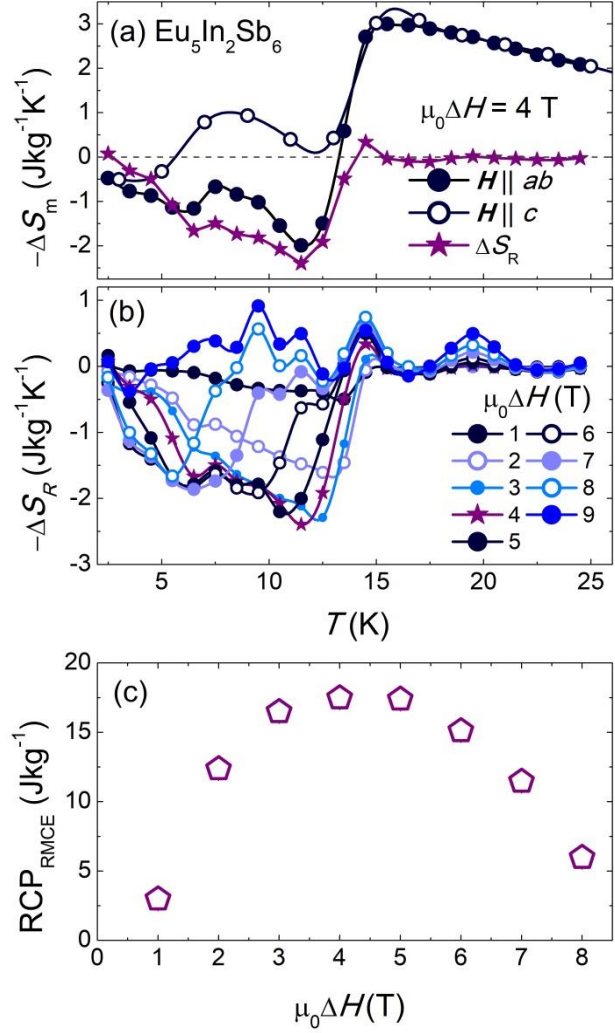

Fig. 6. Rotational magnetocaloric effect (RMCE) in single-crystalline $Eu_5In_2Sb_6$. (a) The magnetic entropy changes as functions of temperature for $\mu_0\Delta H$ = 4 T, determined for the ***H*** || *ab* and ***H*** || *c* orientations, together with the resulting rotational magnetic entropy change. (b) The rotational magnetic entropy change $\Delta S_R$ evaluated for a few values of the magnetic field change. Solid lines are guides for the eyes. (c) The relative cooling power for the rotational magnetocaloric effect plotted as a function of the magnetic field change.

As it was communicated before [4,7–9], the magnetic phase transitions in $Eu_5In_2Sb_6$ are very well visible in the temperature dependence of the specific heat. The results of our heat capacity measurements are shown in Figure 7(a). In zero magnetic field, two lambda-type anomalies in $C_p(T)$, occurring at $T_{N1}$ = 14.0(1) K and $T_{N2}$ = 7.2(1) K, manifest second-order character of the AFM phase transitions. Both anomalies are sensitive to the external magnetic field. With increasing field, they move towards lower temperatures and gradually smear out, in a manner typical for antiferromagnets. At higher temperatures, $C_p(T)$ has a sigmoid-like shape and tends to saturate at a value that corresponds to the classical Dulong-Petit limit ($3n$R = 324 J mol$^{-1}$ K$^{-1}$, where *n* is the number of atoms in the formula unit, and R stands for the gas constant).

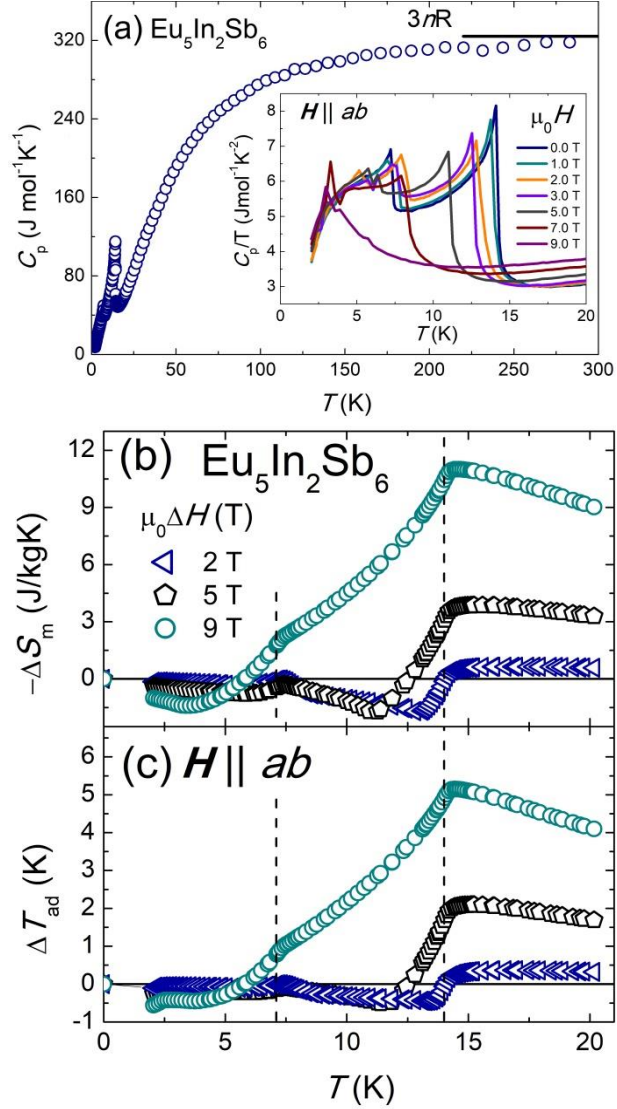

Fig. 7. (a) Temperature dependence of the specific heat of $Eu_5In_2Sb_6$. The solid line represents the Dulong-Petit limit. Inset: low-temperature variations of the specific heat over temperature ratio measured in different magnetic fields applied in the crystallographic *ab*-plane; (b) Temperature variations of the magnetic entropy change calculated from the specific heat data for different values of the magnetic field change; (c) Temperature dependencies of the adiabatic temperature change, calculated as in panel (b). The vertical dashed lines mark the positions of the magnetic phase transition observed in the magnetic measurements.

The specific heat data obtained for $Eu_5In_2Sb_6$ were used to calculate the magnetic entropy change $\Delta S_m$ according to the formula [18]:

$$\Delta S_m(T)_{\mu_0 \Delta H} = \int_0^T \frac{C_p(T,\mu_0 H) - C_p(T,0)}{T} dT = [S_m(T,\mu_0 \Delta H) - S_m(T,0)]. \quad (6)$$

The results of these calculations, shown in Figure 7(b), agree rather well with $\Delta S_m$ afore-determined from the magnetization data. Although, the maximum $\Delta S_m$ values are slightly smaller than those determined from $M(H, T)$, e.g., for $\mu_0 \Delta H = 5$ T, the maximum value of $\Delta S_m$ obtained from $C_p(T)$ equals 3.9(1) J kg$^{-1}$ K$^{-1}$, while that value obtained from $M(H, T)$ amounts to 4.7(1) J kg$^{-1}$ K$^{-1}$. The differences may be due to troublesome determination of the specific heat at $T = 0$ K, which always involves a fairly large uncertainty.

Another important parameter for assessing MCE is the adiabatic temperature change $\Delta T_{ad}$, that can be determined using the equation [18]:

$$\Delta T_{ad} (\Delta H, T) = [T(H, S) - T(0, S)] . \qquad (7)$$

The results obtained for Eu$_5$In$_2$Sb$_6$ are shown in Fig. 7(c). The maximum value of $\Delta T_{ad}$ was found near $T_{N1}$, and it equals 0.4(1) K, 2.1(1) K, and 5.1(1) K for the magnetic field change of 2, 5, and 9 T, respectively.

The Table 1 lists the magnetocaloric parameters value obtained for the classical magnetocaloric effect in Eu$_5$In$_2$Sb$_6$. These values have been compared with previously reported values for other Eu-based materials. The results indicate that the parameter values for Eu$_5$In$_2$Sb$_6$ are not particularly high. However, they are comparable to the parameter values for other Eu-containing multicomponent compounds that are being considered for use in cryogenic magnetic refrigerators. It is possible that by modifying the chemical composition appropriately, the magnetocaloric parameters can be increased.

Table 1. Conventional MCE in several Eu-based intermetallics for the magnetic field change $\mu_0\Delta H = 5$ T. $T_{ord}$ - magnetic ordering temperature, $|\Delta S_m|$ - absolute maximum value of the entropy change, RCP - relative cooling power, $\Delta T_{ad}$ - adiabatic temperature change, TEC(5) - temperature averaged entropy change at 5 K temperature lift.

| Compound | | $T_{ord}$ (K) | $|\Delta S_m^{MAX}|$ (J kg$^{-1}$ K$^{-1}$) | RCP (J kg$^{-1}$) | $\Delta T_{ad}$ (K) | TEC(5) (J kg$^{-1}$ K$^{-1}$) | Ref. |
|---|---|---|---|---|---|---|---|
| Eu$_5$In$_2$Sb$_6$ | $H \parallel ab$ | 14.0/7.1 | 4.7(1) | 81(1) | 2.1(1) | 4.5(1) | This work |
| single crystal | $H \parallel c$ | | 4.7(1) | 81(1) | - | 4.5(1) | |
| EuS | | 18 | 37 | 782 | 10.4 | - | [25] |
| EuSe | | 4.6 | 37.5 | 580 | - | - | [14] |
| EuO | | 69 | 17.5 | 665 | 6.8 | - | [26] |
| Eu$_2$In | | 55 | 34 | 408 | 5* | - | [27] |
| EuCu$_5$In | | 12 | 4.8 | 55.2 | 3.5 | - | [28] |
| EuAu$_5$In | | 13 | 6.4 | 98.4 | 5 | - | [28] |
| Eu$_2$CuSi$_3$ | | 40 | 8.1 | 250 | - | - | [29] |
| EuAuZn | | 52 | 9.1 | 318 | 3.8 | - | [30] |
| EuAuGe | | 33 | 7.6 | 358 | - | - | [31] |
| EuAgZn | | 29 | 14.9 | 409 | - | 4.00** | [32] |
| EuAgCd | | 27 | 13.5 | 321 | - | 4.09** | [32] |
| EuPtZn | | 20 | 15.3 | 280 | - | 3.96** | [32] |
| EuAuCd | | 22 | 13.7 | 311 | - | 4.44** | [32] |
| Eu$_4$PdMg | | 150 | 5.5 | 977 | - | - | [33] |
| EuRhAl$_4$Si$_2$ | | 11.7 | 17.26 | 112 | - | - | [34] |
| EuAl$_4$ | | 15.8 | 21.1 | 282 | - | 20.1 | [35] |
| EuAl$_{3.35}$Si$_{0.65}$ | | 13 | 23.3 | 397 | - | 22.7 | [35] |
| EuAl$_{3.82}$Cu$_{0.18}$ | | 16 | 26.7 | 397 | - | 26.1 | [35] |
| EuIn$_2$As$_2$ | $H \parallel ab$ | 16.1 | ~13 | ~270 | - | - | [36] |
| single crystal | $H \parallel c$ | | ~10 | ~200 | - | - | |
| EuFe$_2$As$_2$ | $H \parallel ab$ | 19 | ~14.6 | 317 | - | - | [37] |
| single crystal | $H \parallel c$ | | ~10.3 | 233 | - | - | |
| EuIr$_2$P$_2$ | $H \parallel ab$ | 5 | - | - | - | - | [38] |
| single crystal | $H \parallel c$ | | 2.5 | - | - | - | |
| EuAu$_3$Al$_2$ | | 16.5 | 4.8 | 82 | - | - | [39] |
| EuTiO$_3$ | | 5.6 | 44.4 | 353 | - | - | [40] |
| Eu$_3$O$_4$ | | 5.3 | 12.7 | 159 | 7 | - | [41] |

* for $\mu_0\Delta H = 2$ T.

** TEC(3) for $\mu_0\Delta H$ = 1 T.

## SUMMARY

The single crystals of $Eu_5In_2Sb_6$ exhibit a noticeable magnetocaloric effect over a wide temperature range. It demonstrates two consecutive magnetic phase transitions and displays both normal and inverse MCE, along with rotational MCE due to its significant magnetic anisotropy. The conventional MCE shows a maximum isothermal entropy change of about -4.7(1) J kg$^{-1}$ K$^{-1}$ at 15(1) K for a 5 T magnetic field change, independent of field direction. The relative cooling power (RCP) is consistent for both directions, at 81(1) J kg$^{-1}$ for a 5 T magnetic field change. The rotational MCE is most notable at around 11 K, with a maximum $\Delta S_R$ value of 2.4(1) J kg$^{-1}$ K$^{-1}$ and RCP = 17.5(6) J kg$^{-1}$ for a 4 T magnetic field change. While the MCE parameters for $Eu_5In_2Sb_6$ are not the highest, they are comparable to those of other multi-component Eu-based intermetallics that have been identified as potential candidates for magnetocaloric materials.

## ACKNOWLEDGEMENTS

This study was supported by the National Science Centre (Poland) under grant 2021/41/B/ST3/01141.